%% file: letter.tex
\documentclass[10pt,aps,prl,twocolumn,preprintnumbers,nofootinbib]{revtex4-2}
\usepackage{amsmath,amssymb,mathtools}
\usepackage{graphicx}
\usepackage{booktabs,cellspace}
\usepackage{hyperref}
\usepackage{xspace}

\newcommand{\as}{\ensuremath{\alpha_s}\xspace}

\newcommand{\pb}{\ensuremath{\mathrm{pb}}\xspace}

\newcommand{\GeV}{\ensuremath{\mathrm{GeV}}\xspace}
\newcommand{\TeV}{\ensuremath{\mathrm{TeV}}\xspace}
\newcommand{\mll}{\ensuremath{{m_{{\ell\ell}}}}\xspace}
\newcommand{\pt}{\ensuremath{{p_T^{{\ell\ell}}}}\xspace}
\newcommand{\ptcut}{\ensuremath{{p_T^{\mathrm{cut}}}}\xspace}
\newcommand{\ptcutmin}{\ensuremath{{p_{T,\,\text{min}}^{\mathrm{cut}}}}\xspace}
\newcommand{\ptoff}{\ensuremath{{p_T^{\mathrm{off}}}}\xspace}
\newcommand{\mur}{\ensuremath{{\mu_{\mathrm{R}}}}\xspace}
\newcommand{\muf}{\ensuremath{{\mu_{\mathrm{F}}}}\xspace}
\newcommand{\ntlo}{\text{N${}^3$LO}\xspace}
\newcommand{\ntll}{\text{N${}^3$LL}\xspace}
\newcommand{\nnlo}{\text{NNLO}\xspace}
\newcommand{\dy}{\text{Drell--Yan}\xspace}
\newcommand{\rd}{\ensuremath{{\mathrm{d}}}\xspace}
\newcommand{\RadISH}{\texttt{RadISH}\xspace}
\newcommand{\NNLOJET}{\texttt{NNLOJET}\xspace}

\begin{document}

\title{Third order fiducial predictions for Drell--Yan at the LHC}

\preprint{CERN-TH-2022-023, IPPP/22/08, LAPTH-009/22, KA-TP-03-2022, P3H-22-022, ZU-TH 07/22}

\author{%
  Xuan Chen$^{1,2}$,
  Thomas Gehrmann$^3$,
  Nigel Glover$^4$,
  Alexander Huss$^5$,
  Pier Francesco Monni$^5$,
  Emanuele Re$^{6,7}$,
  Luca Rottoli$^3$,
  Paolo Torrielli$^8$
\bigskip}

\affiliation{$^1$ Institute for Theoretical Physics, Karlsruhe Institute of Technology, 76131 Karlsruhe, Germany}
\affiliation{$^2$ Institute for Astroparticle Physics, Karlsruhe Institute of Technology, 76344 Eggenstein-Leopoldshafen, Germany}
\affiliation{$^3$ Department of Physics, University of Z\"urich, CH-8057 Z\"urich, Switzerland}
\affiliation{$^4$ Institute for Particle Physics Phenomenology, Physics Department, Durham University, Durham DH1 3LE, UK}
\affiliation{$^5$ CERN, Theoretical Physics Department, CH-1211 Geneva 23, Switzerland}
\affiliation{$^6$ Dipartimento di Fisica G. Occhialini,
  U2, Universit\`a degli Studi di Milano-Bicocca and INFN,
  Sezione di Milano-Bicocca,
  Piazza della Scienza, 3, 20126 Milano, Italy}
\affiliation{$^7$ LAPTh, Universit\'e Grenoble Alpes, Universit\'e Savoie Mont Blanc, CNRS, F-74940 Annecy, France}
\affiliation{$^8$ Dipartimento di Fisica and Arnold-Regge Center,
  Universit\`a di Torino and INFN, Sezione di Torino,
  Via P. Giuria 1, I-10125, Turin, Italy}

\begin{abstract}
  The Drell--Yan process at hadron colliders is a fundamental
  benchmark for the study of strong interactions and the
  extraction of electro-weak parameters. The outstanding precision of
  the LHC demands very accurate theoretical predictions with a full
  account of fiducial experimental cuts.
  In this letter we present a state-of-the-art calculation of the
  fiducial cross section and of differential distributions for this
  process at third order in the strict fixed-order expansion in
  the strong coupling, as well as including the all-order resummation of
  logarithmic corrections.
  Together with these results, we present a detailed study of the
  subtraction technique used to carry out the calculation for
  different sets of experimental cuts, as well as of the sensitivity
  of the fiducial cross section to infrared physics.
  We find that residual theory uncertainties are reduced to the
  percent level and that the robustness of the 
  predictions can be improved by a suitable adjustment of fiducial cuts.\sloppy
  \end{abstract}

\maketitle

\paragraph{Introduction.---}%
The fine understanding of Quantum
Chromodynamics (QCD) demanded by the physics programme of the Large
Hadron Collider (LHC) has led to the impressive development of new
computational techniques to achieve precise predictions for
hadronic scattering reactions.
Among these, the production of a lepton pair (the \dy
process)~\cite{Drell:1970wh} arguably constitutes the most important
standard candle at hadron colliders. The precise data collected at the
LHC enables a broad spectrum of high-profile applications to different
areas of particle physics, such as the extraction of Standard Model
(SM)
parameters~\cite{ATLAS:2017rzl,Ball:2018iqk,Bertacchi:2019mqn,Bagnaschi:2019mzi,LHCb:2021bjt} and of the
parton densities of the proton~\cite{Boughezal:2017nla}, and the
exploration of beyond the Standard Model
scenarios~\cite{CMS:2021ctt,ATLAS:2021mla}.
At present, the theoretical description of this important reaction reaches
the highest-yet level of perturbative accuracy. Fixed-order perturbative predictions in
QCD, obtained as an expansion in the strong coupling \as, are
known up to third order beyond the Born approximation, i.e.\
next-to-next-to-next-to-leading order (\ntlo), for the \dy cross
section and rapidity distribution calculated inclusively over the
phase space of QCD
radiation~\cite{Duhr:2020sdp,Duhr:2020seh,Duhr:2021vwj,Chen:2021vtu}.
Moreover, next-to-next-to-leading order (\nnlo) corrections for the
production of a \dy pair in association with one QCD jet have been
computed in
Refs.~\cite{Boughezal:2015dva,Gehrmann-DeRidder:2015wbt,Boughezal:2015ded,Boughezal:2016dtm,Boughezal:2016isb,Gehrmann-DeRidder:2016cdi,Gehrmann-DeRidder:2016jns,Gauld:2017tww,Gehrmann-DeRidder:2017mvr,Gauld:2021pkr}.
Similarly, electro-weak (EW) corrections are known up to
next-to-leading order
(NLO)~\cite{Dittmaier:2001ay,Baur:2001ze,Baur:2004ig,Arbuzov:2005dd,Zykunov:2005tc,Zykunov:2006yb,CarloniCalame:2006zq,CarloniCalame:2007cd,Arbuzov:2007db,Dittmaier:2009cr}
and mixed QCD--EW at
\nnlo~\cite{Dittmaier:2014qza,Dittmaier:2015rxo,Bonciani:2016wya,deFlorian:2018wcj,Bonciani:2019nuy,Delto:2019ewv,Cieri:2020ikq,Bonciani:2020tvf,Buccioni:2020cfi,Behring:2020cqi,Buonocore:2021rxx,Bonciani:2021zzf}.
The description of kinematical distributions sensitive to the emission
of soft and/or collinear QCD radiation features large logarithms of the transverse
momentum of the \dy pair.
 The presence of such logarithmic-enhanced terms at all orders in perturbation theory spoils a fixed-order description and
demands in addition the resummation of
radiative corrections at all orders in the strong
coupling~\cite{Parisi:1979se,Collins:1984kg,Balazs:1997xd,Catani:2000vq,Bozzi:2010xn,Becher:2010tm,Bizon:2017rah}. Currently
such calculations have been performed up to
next-to-next-to-next-to-leading logarithmic (\ntll) accuracy~\cite{Bizon:2018foh,Becher:2019bnm,Bizon:2019zgf,Ebert:2020dfc,Becher:2020ugp}, also including
the analytic constant terms up to $\mathcal{O}(\as^3)$ in
Refs.~\cite{Camarda:2021ict,Re:2021con,Ju:2021lah}, enabled by the
perturbative ingredients from
Refs.~\cite{Gehrmann:2010ue,Catani:2012qa,Gehrmann:2014yya,Lubbert:2016rku,
  Echevarria:2016scs,Li:2016ctv,Vladimirov:2016dll,Moch:2017uml,Moch:2018wjh,Lee:2019zop,Henn:2019swt,
  Bruser:2019auj,Henn:2019rmi,vonManteuffel:2020vjv,Luo:2019szz,Ebert:2020yqt,Luo:2020epw}.
Additional sources of logarithmic corrections have been
considered~\cite{Laenen:2000ij,Berge:2005rv,Marzani:2015oyb,Lustermans:2016nvk,Pietrulewicz:2017gxc,Cieri:2018sfk,Rabemananjara:2020rvw},
and found to have a moderate numerical impact.
The modelling of effects beyond collinear factorisation, relevant for
low-mass \dy production, has also been studied (see e.g.\
Refs.~\cite{Scimemi:2019cmh,Bertone:2019nxa,Bacchetta:2019sam,Hautmann:2020cyp,BermudezMartinez:2020tys,Martinez:2021chk}).
Finally, the high accuracy of the experimental measurements of this process makes it an
ideal laboratory for the development of state-of-the-art event
generators~\cite{Hoche:2014uhw,Karlberg:2014qua,Alioli:2015toa,Monni:2019whf,Monni:2020nks,Alioli:2021qbf}.

Despite this outstanding progress, the accurate description of
experimental data is challenged by the presence of fiducial selection
cuts in the measurements, whose inclusion in theoretical calculations
can potentially compromise the stability of the perturbative
expansion~\cite{Klasen:1995xe,Harris:1997hz,Frixione:1997ks,Salam:2021tbm}.
An initial estimate of the \ntlo \dy cross section with an account of
experimental cuts was presented in
Refs.~\cite{Camarda:2021ict,Camarda:2021jsw} using the
\mbox{$q_T$-subtraction formalism}~\cite{Catani:2007vq}, albeit
without a complete assessment of the theoretical and methodological
uncertainties. The conclusions of the above study are discussed in
detail in Appendix.

In this letter, we present state-of-the-art predictions both for the
fiducial \dy cross section and for differential distributions of the
final-state leptons. We exploit this calculation to carry out, for the
first time, a thorough study of the robustness of these theory
predictions in the presence of different sets of fiducial cuts.
We also present a detailed analysis of the reliability of the
computational method adopted, and show that reaching a robust control
over the involved systematic uncertainties requires an excellent
stability of the numerical calculation in deep infrared kinematic
regimes.

\paragraph{Methodology.---}%
The starting point of our calculation
for the production cross section $\rd\sigma_{\text{DY}}$
of a \dy lepton pair, differential in its phase space and in the
pair's transverse momentum \pt, is the formula:
\begin{equation}
  \label{eq:master}
  \rd\sigma_{\text{DY}}^{\ntlo+\ntll} \equiv
  \rd\sigma_{\text{DY}}^{\ntll} +
  \rd\sigma_{\text{DY+jet}}^{\nnlo} -
  \big[\rd\sigma_{\text{DY}}^{\ntll}\big]_{{\mathcal{O}(\as^3)}}
\end{equation}
where $\rd\sigma_{\text{DY}}^{\ntll}$ represents the \ntll resummed
\pt distribution obtained in Ref.~\cite{Re:2021con} with the
computer code
\RadISH~\cite{Monni:2016ktx,Bizon:2017rah,Monni:2019yyr},
including the analytic constant terms up to $\mathcal{O}(\as^3)$;
the quantity
$\big[\rd\sigma_{\text{DY}}^{\ntll}\big]_{{\mathcal{O}(\as^3)}}$
is its expansion up to third order in \as, and
$\rd\sigma_{\text{DY+jet}}^{\nnlo} $ is the differential \pt
distribution at \nnlo~(i.e. $\mathcal O(\as^3)$), obtained with the \NNLOJET
code~\cite{Gehrmann-DeRidder:2015wbt,Gehrmann-DeRidder:2016cdi,Gehrmann-DeRidder:2016jns}.
Eq.~\eqref{eq:master} is finite in the limit $\pt\to 0$: by
integrating it inclusively over \pt one can obtain predictions
differential in the leptonic phase space at \ntlo{}+\ntll perturbative
accuracy, allowing for the inclusion of fiducial cuts.
An important challenge in the evaluation of the integral of
Eq.~\eqref{eq:master} over \pt is given by the fact that both
$\rd\sigma_{\text{DY+jet}}^{\nnlo} $ and
$\big[\rd\sigma_{\text{DY}}^{\ntll}\big]_{{\mathcal{O}(\as^3)}}$
diverge logarithmically in the limit $\pt\to 0$, and only their
difference is finite since the large logarithmically divergent terms
present in $\rd\sigma_{\text{DY+jet}}^{\nnlo} $ are exactly matched by those contained in $\big[\rd\sigma_{\text{DY}}^{\ntll}\big]_{{\mathcal{O}(\as^3)}}$.
Guaranteeing the cancellation of such divergences requires high
numerical precision in the \nnlo distribution
$\rd\sigma_{\text{DY+jet}}^{\nnlo} $ down to very small values of
\pt.
Setting
$\rd\sigma_{\text{DY+jet}}^{\nnlo}
-\big[\rd\sigma_{\text{DY}}^{\ntll}\big]_{{\mathcal{O}(\as^3)}}= 0$
for $\pt\leq\ptcut$ introduces a slicing error of order
$\mathcal{O}((\ptcut/\mll)^n)$. If one integrates inclusively over
the leptonic phase space one has
$n=2$, while the presence of fiducial cuts in general leads to the
appearance of linear terms with
$n=1$~\cite{Grazzini:2017mhc,Ebert:2019zkb,Alekhin:2021xcu,Salam:2021tbm}. 
Starting from order $\as^2$, the corrections are further enhanced by
logarithms of $\ptcut$. The presence of these corrections introduces a
systematic uncertainty which can be controlled by reducing the value of $\ptcut$
to a sufficiently small value. This procedure is computationally demanding especially
in the presence of linear corrections, due to the smaller value of $\ptcut$ required to achieve
the independence of the results of the slicing parameter.
Such linear corrections can be resummed at all orders in
Eq.~\eqref{eq:master}~\cite{Ebert:2020dfc} by applying a
simple recoil prescription~\cite{Catani:2015vma} to
$\rd\sigma_{\text{DY}}^{\ntll}$, and their inclusion
would in principle allow for a larger \ptcut in the
calculation.
These effects are accounted for in Eq.~\eqref{eq:master}, as discussed
in Ref.~\cite{Re:2021con}. As a consequence, our \ntlo{}+\ntll
fiducial predictions obtained by integrating Eq.~\eqref{eq:master} are
only affected by a slicing error of order
$\mathcal{O}((\ptcut/\mll)^2)$.

The perturbative expansion of the \ntlo{}+\ntll fiducial cross
section to third order in \as leads to the \ntlo prediction
as obtained according to the $q_T$-subtraction
formalism~\cite{Catani:2007vq}.
In this case, the outlined procedure to include linear power
corrections below \ptcut in the \ntlo computation is analogous to
that of Refs.~\cite{Buonocore:2021tke,Camarda:2021jsw}.
Since the fiducial cross section can be computed up to \nnlo using the
\NNLOJET code, which implements a subtraction
technique~\cite{Gehrmann-DeRidder:2005btv,Currie:2013vh} that does not
require the introduction of a slicing parameter, in the fixed-order
results quoted in this letter we apply the above procedure only to the
computation of the \ntlo correction, while retaining the
$\ptcut$-independent result up to \nnlo.
This effectively suppresses the slicing error in our
fiducial \ntlo cross section to
$\mathcal{O}(\as^3\, (\ptcut/\mll)^2)$.

\begin{figure}[t!]
  \centering
  \includegraphics[width=\linewidth]{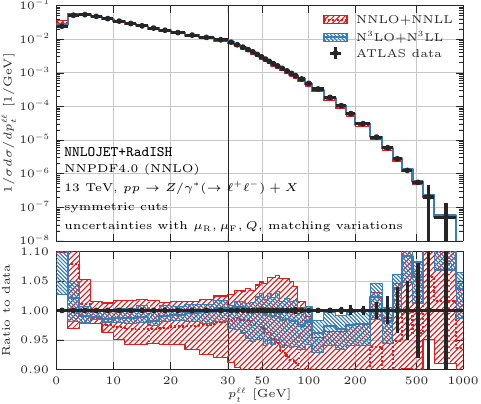}
  \caption{\label{fig:pt-spectrum} Fiducial \pt distribution at
    \ntlo{}+\ntll (blue, solid) and NNLO+NNLL (red, dotted) compared
    to \texttt{ATLAS} data from Ref.~\cite{ATLAS:2019zci}. The binning is linear up to $30\,\GeV$
     and logarithmic above.
    }
\end{figure}

\begin{table*}[t!]
\centering
\begin{tabular}{@{} c !{\hspace{.5em}} >{\enspace}Sl>{\enspace}Sl !{\hspace{.5em}} >{\enspace}Sl>{\enspace}Sl @{}}
  \toprule
  Order &
  \multicolumn{2}{c}{$\sigma\; [\pb]$  Symmetric cuts} &
  \multicolumn{2}{c}{$\sigma\; [\pb]$  Product cuts}
  \\[1ex]
  $k$ &
  \multicolumn{1}{l}{ N${}^k$LO } &
  \multicolumn{1}{l}{ N${}^k$LO+N${}^k$LL } &
  \multicolumn{1}{l}{ N${}^k$LO } &
  \multicolumn{1}{l}{ N${}^k$LO+N${}^k$LL }
  \\
  \cmidrule(r{.5em}){1-1} \cmidrule(r{.5em}){2-3} \cmidrule{4-5}
  $0$                          &
  $721.16^{+12.2\%}_{-13.2\%}$ & --- &
  $721.16^{+12.2\%}_{-13.2\%}$ & ---
  \\
  $1$                           &
  $742.80(1)^{+2.7\%}_{-3.9\%}$ & $748.58(3)^{+3.1\%}_{-10.2\%}$ &
  $832.22(1)^{+2.7\%}_{-4.5\%}$ & $831.91(2)^{+2.7\%}_{-10.4\%}$
  \\
  $2$                             &
  $741.59(8)^{+0.42\%}_{-0.71\%}$ & $740.75(5)^{+1.15\%}_{-2.66\%}$ &
  $831.32(3)^{+0.59\%}_{-0.96\%}$ & $830.98(4)^{+0.74\%}_{-2.73\%}$
  \\
  $3$                                      &
  $722.9(1.1)^{+0.68\%}_{-1.09\%} \pm 0.9$ & $726.2(1.1)^{+1.07\%}_{-0.77\%}$ &
  $816.8(1.1)^{+0.45\%}_{-0.73\%} \pm 0.8$ & $816.6(1.1)^{+0.87\%}_{-0.69\%}$
  \\
  \bottomrule
\end{tabular}
\caption{\label{tab:xs}Fiducial cross sections for the symmetric Eq.~\eqref{eq:atlas}
  and product Eq.~\eqref{eq:gavin} cuts both at fixed perturbative order
  and including all-order resummation. We report the theoretical uncertainty in percent and, in parentheses, the absolute value of the statistical uncertainty. The latter applies to the last significant figures displayed.
At N$^3$LO we also separately indicate the slicing error, in absolute value. See the main text for details.}
\end{table*}

In general, the presence of linear fiducial power corrections indicates an
arguably undesirable sensitivity of the fiducial cross section to the
infrared region in which QCD radiation has small transverse
momentum, which compromises the stability of the
perturbative series~\cite{Salam:2021tbm}.
These issues can be avoided by modifying the definition of the fiducial cuts in
such a way that the scaling of the power corrections be quadratic
across most of the leptonic phase space.
In the following we present a calculation of Eq.~\eqref{eq:master} and
of the fiducial cross section both for the standard (\emph{symmetric})
cuts adopted by LHC
experiments~\cite{ATLAS:2019zci,CMS:2019raw}, where the same 
cut is imposed on transverse momentum of the final state leptons, as well as for the
modified (\emph{product}) cuts proposed in
Ref.~\cite{Salam:2021tbm},  where a cut is instead 
imposed on the product of the transverse momenta of the final state leptons.
This state-of-the-art calculation allows us to assess precisely the
effect of different types of fiducial cuts on the theoretical prediction
for the cross section, as well as on the performance of the computational
approach adopted here.

\paragraph{Results.---}%

We consider proton--proton collisions at a
centre-of-mass energy $\sqrt{s}=13\,\TeV$. We adopt the
\texttt{NNPDF4.0} parton densities~\cite{Ball:2021leu} at NNLO with
$\as(M_Z)=0.118$, whose scale evolution is
performed with \texttt{LHAPDF}~\cite{Buckley:2014ana} and
\texttt{Hoppet}~\cite{Salam:2008qg}, correctly accounting for heavy quark
thresholds.
We adopt the $G_\mu$ scheme with the following EW parameters
taken from the PDG~\cite{ParticleDataGroup:2018ovx}: $M_Z = 91.1876\,\GeV$,
$M_W = 80.379\,\GeV$, $\Gamma_Z = 2.4952\,\GeV$,
$\Gamma_W = 2.085\,\GeV$, and
$G_F=1.1663787\times 10^{-5}\,\GeV^{-2}$.
We consider two fiducial volumes, in both of which the leptonic
invariant-mass window is $66\,\GeV < \, \mll \,< 116\,\GeV$ and
the lepton rapidities are confined to $|\eta^{\ell^\pm}| < 2.5$.
The transverse momentum of the two leptons is constrained as
\begin{subequations}
\begin{align}
  \label{eq:atlas}
  &\text{Symmetric cuts~\cite{ATLAS:2019zci}:}
  &|\vec{p}_{T}^{~\ell^\pm}|  &> 27\,\GeV\,,\\
  \label{eq:gavin}
  &\text{Product cuts~\cite{Salam:2021tbm}:}
  &\sqrt{|\vec{p}_{T}^{~\ell^+}|\,|\vec{p}_{T}^{~\ell^-}|} &> 27\,\GeV\,,
  \notag\\
  && \min\{|\vec{p}_{T}^{~\ell^\pm}|\} &> 20\,\GeV\,.
\end{align}
\end{subequations}

The central factorisation and renormalisation scales are chosen to be
$\mur = \muf = \sqrt{\mll^2+\pt^2}$ and the
central resummation scale is set to $Q=\mll/2$.
In the results presented below, the theoretical uncertainty is
estimated by varying the \mur and \muf scales by a factor of two
about their central value, while keeping
$1/2 \leq \mur/\muf \leq 2$.
In addition, for the resummed results, for central $\mur = \muf$
scales we vary $Q$ by a factor of two around its central
value. Moreover, a matching-scheme uncertainty is estimated
by including the full scale variation of the additive matching
scheme of Ref.~\cite{Re:2021con} (27 variations that
comprise the one of the central matching scale $v_0$ introduced
in Eq.~(5.2) of that article).
The final uncertainty is obtained as the envelope of all the above
variations, corresponding to 7 and 36 curves for the fixed-order
and resummed computations, respectively.
We present results for the central member of the 
\texttt{NNPDF4.0} set.
In the fiducial cross sections quoted below at \ntlo and
\ntlo{}+\ntll, we do not consider the uncertainty related to the
missing \ntlo parton distributions, which are currently unavailable.

In Fig.~\ref{fig:pt-spectrum}, we start by showing the
transverse-momentum distribution of the \dy lepton pair in the fiducial
volume Eq.~\eqref{eq:atlas}, obtained with Eq.~\eqref{eq:master}, compared
to experimental data~\cite{ATLAS:2019zci}.
%
%
In the figure we label the distributions by the perturbative accuracy
of their inclusive integral over \pt.
Our state-of-the-art \ntlo{}+\ntll prediction provides an excellent
description of the data across the spectrum, with the exception of the
first bin at small \pt which is susceptible to non-perturbative
corrections not included in our calculation.
We point out that the term
$\rd\sigma_{\text{DY+jet}}^{\nnlo}
-\big[\rd\sigma_{\text{DY}}^{\ntll}\big]_{{\mathcal{O}(\as^3)}}$ in
Eq.~\eqref{eq:master} gives a non-negligible contribution even for
$\pt \leq 15\,\GeV$.
The residual theoretical uncertainty in the intermediate \pt region
is at the few-percent level, and it increases to about $5\%$ for
$\pt\gtrsim 50\,\GeV$.
A more accurate description of the large-\pt region requires the
inclusion of EW corrections, which we neglect in our
calculation.

We now consider the fiducial cross section with symmetric
cuts.
In order to gain control over the slicing systematic error, we choose
\ptcut as low as $0.81\,\GeV$.
In the first column of Tab.~\ref{tab:xs}, denoted as N${}^k$LO,
we show the fixed-order results to $\mathcal{O}(\as^k)$.
The second column of Tab.~\ref{tab:xs} displays the result obtained
including resummation effects.
In the fixed-order case, the theoretical uncertainty at \ntlo, estimated as discussed above, is
supplemented with an estimate of the slicing uncertainty obtained by
varying \ptcut in the range $[0.45,\,1.48]\,\GeV$ and taking the average difference from the
result with $\ptcut=0.81\,\GeV$.
In the resummed case, we quote the total theoretical uncertainty
including also the matching scheme variation.
In both cases the statistical uncertainty is reported in parentheses.

We observe that the new \ntlo corrections decrease the fiducial
cross section by about $2.5\%$, and the final
prediction at \ntlo has larger theoretical errors than the \nnlo
counterpart, whose uncertainty band does not capture the \ntlo
central value.
This indicates a poor convergence of the fixed-order perturbative
series for this process, which is consistent with what has been
observed in the inclusive case in
Refs.~\cite{Duhr:2020seh,Duhr:2020sdp,Duhr:2021vwj}.
In the resummed case, the theoretical uncertainty is more reliable and within errors
the convergence of the perturbative series is improved.
The presence of linear power corrections is also responsible for the
moderate difference between the fixed-order and the resummed
prediction for the symmetric cuts, which as
previously discussed indicates a sensitivity of the cross section to the infrared region of
small \pt.
This ultimately worsens further the perturbative convergence of the
fixed-order series thereby challenging the perspectives to reach
percent-accurate theoretical predictions within symmetric cuts.

A possible solution to this problem~\cite{Salam:2021tbm} is to
slightly modify the definition of the fiducial cuts as in
Eq.~\eqref{eq:gavin} in order to reduce such a sensitivity to infrared
physics.
We present for the first time theoretical predictions up to \ntlo and
\ntlo{}+\ntll for this set of cuts, reported in the third and fourth
column of Tab.~\ref{tab:xs}.
The relative difference between the fixed-order and resummed
calculations for the fiducial cross section never exceeds
$0.04\%$, which indicates that the predictions with product cuts can
be computed accurately with fixed-order perturbation theory.
Nevertheless, we still observe a more reliable estimate of the
theoretical uncertainties when resummation is included.

\begin{figure}[t!]
  \centering
  \includegraphics[width=\linewidth]{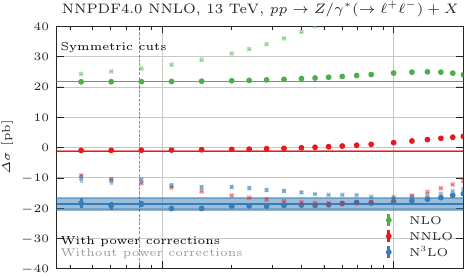} \\
  \includegraphics[width=\linewidth]{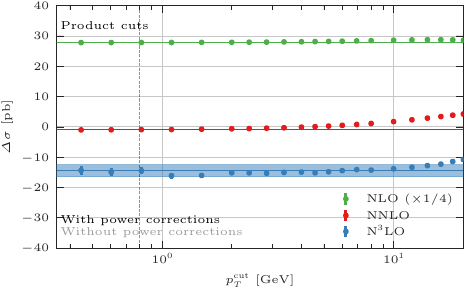}
  \caption{\label{fig:ptcut} Dependence of the extracted N${}^k$LO corrections to the
    fiducial cross sections shown in Tab.~\ref{tab:xs} on the \ptcut
    infrared parameter, both for the symmetric and
    product cuts.  In the latter case, the NLO
    correction has been rescaled by a factor $1/4$.  The dashed
    vertical line indicates our default value $\ptcut=0.81\,\GeV$. The
    blue band is obtained by combining linearly the statistical and
    slicing errors.}
\end{figure}

In order to study the stability of our predictions against variations
of the infrared parameter \ptcut, in Fig.~\ref{fig:ptcut} we show the
dependence of the N${}^k$LO correction (i.e.\ the $\mathcal{O}(\as^k)$
term in the expansion of the fiducial cross section) on $\ptcut$ down
to $\ptcut \simeq 0.4\,\GeV$.
In the case of symmetric cuts Eq.~\eqref{eq:atlas}, we observe that the
inclusion of the linear power corrections is essential to reach a
plateau at small \ptcut, achieving the necessary independence of the
result on the slicing parameter.
We thus obtain an excellent control over the estimate of the slicing
error quoted in Tab.~\ref{tab:xs}.
Furthermore, Fig.~\ref{fig:ptcut} clearly shows that the omission of such
linear corrections leads to an incorrect result for the fiducial
cross section computed with the $q_T$-subtraction method, unless
$\rd\sigma_{\text{DY+jet}}^{\nnlo}$ can be computed precisely down to
$\ptcut \ll 1\,\GeV$.
Conversely, in the case of the product cuts, we
observe a much milder dependence of the N${}^k$LO correction on
\ptcut, and the further inclusion of power corrections does not lead
to any visible difference, consistent with the fact that such
corrections are quadratic in most of the phase
space~\cite{Salam:2021tbm}.
As an additional sanity check, we have repeated the test of
Fig.~\ref{fig:ptcut} for each individual flavour channel contributing
to the \ntlo \dy cross section. The results are collected in the
Appendix, together with a discussion
on alternative approaches to $q_T$ subtraction employing a fitting
procedure~\cite{Billis:2021ecs}, and a comparison to the literature~\cite{Camarda:2021ict,Camarda:2021jsw}.

\begin{figure}[t!]
  \centering
  \includegraphics[width=\linewidth]{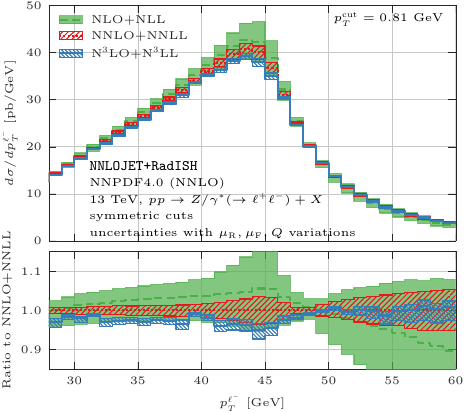}
  \caption{\label{fig:ptlep} Lepton transverse momentum distribution
    up to \ntlo{}+\ntll order in the fiducial phase
    space Eq.~\eqref{eq:atlas}. The labels indicate the order in the
    fiducial cross section.}
\end{figure}

Finally, the computation presented in this letter allows us to
obtain, for the first time, \ntlo{}+\ntll predictions for the
kinematical distributions of the final-state leptons.
A particularly relevant distribution is the leptonic transverse
momentum, which plays a central role in the precise extraction of the
$W$-boson mass at the LHC~\cite{ATLAS:2017rzl,LHCb:2021bjt}.
Fig.~\ref{fig:ptlep} shows the differential distribution of the
negatively charged lepton at three different orders, for our default
value $\ptcut=0.81\,\GeV$. Unlike for the fiducial cross section, the
inclusion of $\pt$ resummation in this observable is crucial to cure
local (integrable) divergences in the spectrum due to the presence of
a Sudakov shoulder~\cite{Catani:1997xc} at
$p_{T}^{\ell^-} \sim \mll/2$.
The figure shows an excellent convergence of the perturbative
prediction, with residual uncertainties at \ntlo{}+\ntll of the order
of a few percent across the entire range.

\paragraph{Conclusions.---}%
In this letter, we have presented state-of-the-art predictions
for the fiducial cross section and differential
distributions in the \dy process at the LHC, through both
\ntlo and \ntlo{}+\ntll in QCD.
These new predictions are obtained through the combination of an
accurate NNLO calculation for the production of a \dy pair in
association with one jet, and the \ntll resummation of logarithmic
corrections arising at small \pt.
The high quality of these results allowed us to carry out a thorough
study of the performance of the computational method adopted, reaching
an excellent control over all systematic uncertainties involved.
We presented predictions for two different definitions of the fiducial
volumes, relying either on symmetric cuts Eq.~\eqref{eq:atlas} on the
transverse momentum of the leptons, or on a recently proposed
product cuts Eq.~\eqref{eq:gavin} which is shown to improve the stability of
the perturbative series.
Our results display residual theoretical uncertainties at the
$\mathcal{O}(1\%)$ level in the fiducial cross section, and at the
few-percent level in differential distributions.
These predictions will play an important role in the comparison of
experimental data with an accurate theoretical description of the \dy
process at the LHC.

\begin{acknowledgments}
\paragraph{Acknowledgments.---}
We are grateful to Luca Buonocore, Massimiliano Grazzini, and Gavin
Salam for discussions and constructive comments on the
manuscript, and to Aude Gehrmann--De Ridder, Tom Morgan, and Duncan Walker for their contributions to the $V+\text{jet}$ process in the \NNLOJET code.
This work has received funding from the Deutsche Forschungsgemeinschaft (DFG, German Research Foundation) under grant 396021762-TRR 257, from the Swiss National Science Foundation (SNF) under contracts
PZ00P2$\_$201878, 200020$\_$188464 and 200020$\_$204200,
from the UK Science and Technology Facilities Council (STFC) through grant ST/T001011/1, from the Italian Ministry of University and Research (MIUR) through grant PRIN 20172LNEEZ,
and from the European Research Council (ERC) under the European Union's Horizon 2020 research and innovation programme grant agreement 101019620 (ERC Advanced Grant TOPUP). 
\end{acknowledgments}

\bibliographystyle{apsrev4-2}
\bibliography{letter}

\input{supplementary_material}

\end{document}

%% file: supplementary_material.tex
\newpage

\onecolumngrid
\newpage
\appendix

\section{Appendix}

In this appendix we provide additional details on the consistency checks of the calculation presented in the letter, as well as on comparisons to other variants of the $q_T$-subtraction method adopted in the literature.

\subsection{Dependence on $\ptcut$ for the individual flavour channels}
\begin{figure}[h!] \includegraphics[width=0.3\linewidth]{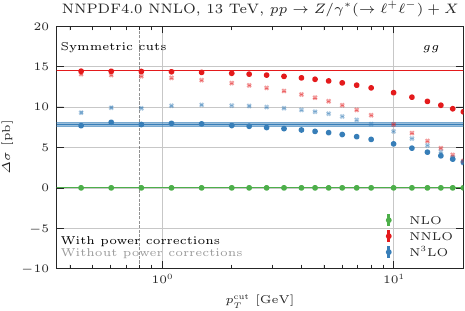}
  \includegraphics[width=0.3\linewidth]{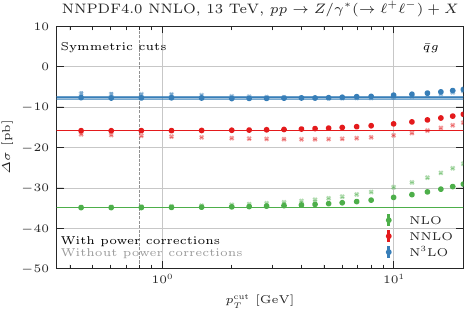}
  \includegraphics[width=0.3\linewidth]{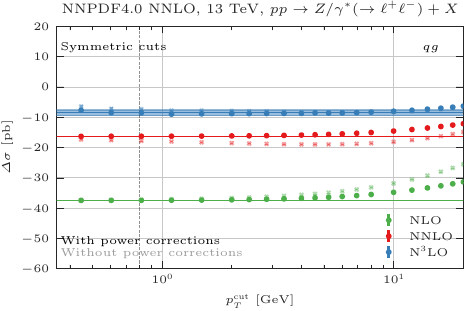}\\
\includegraphics[width=0.3\linewidth]{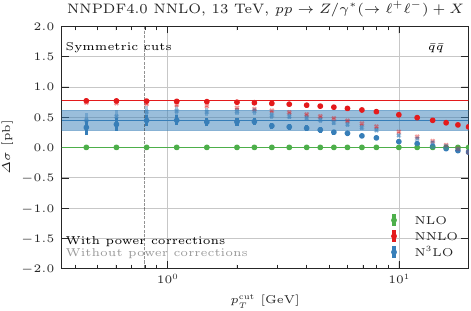}
\includegraphics[width=0.3\linewidth]{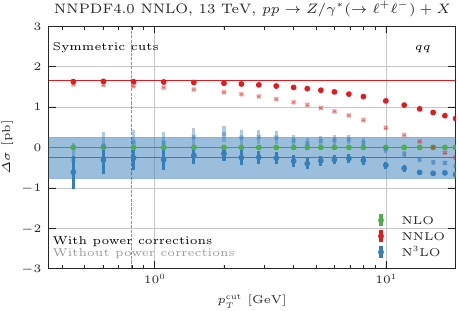}
\includegraphics[width=0.3\linewidth]{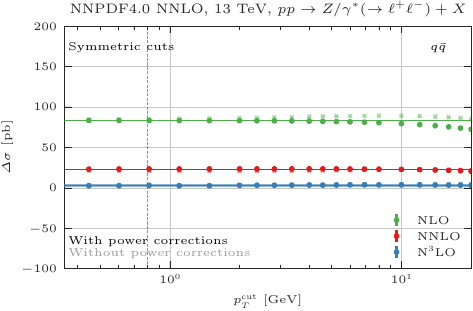}
  \caption{\label{fig:ptcut-channels} Dependence on $\ptcut$ of the extracted N${}^k$LO
    fiducial cross sections for the individual flavour channels. The
    blue band is obtained by combining linearly the statistical and
    slicing errors. }
\end{figure}

Here we provide a breakdown of the results shown in Fig.~2 into the different flavour channels contributing to the \ntlo{} cross section with the fiducial cuts of Eq.~(2a). The results are displayed in Fig.~\ref{fig:ptcut-channels}, where we observe a clear independence of the extracted N$^k$LO corrections on $\ptcut$ for each individual channel. 
An analogous independence of $\ptcut$ is observed with the fiducial cuts of Eq.~(2b), not shown here. 
This constitutes a strong consistency check of our calculation.

\subsection{Comparison to the literature}
\begin{figure}[h!]
  \centering \includegraphics[width=0.6\linewidth]{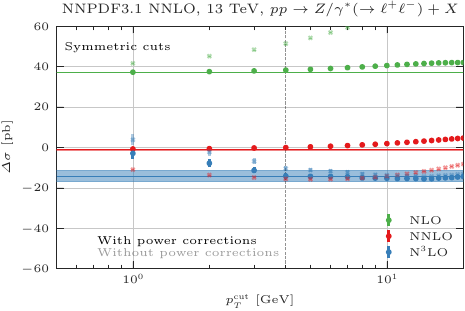}
  \caption{\label{fig:ptcut-suppl} Dependence on $\ptcut$ of the extracted N${}^k$LO
    fiducial cross sections using the data from Ref.~\cite{Bizon:2019zgf}. The
    blue band is obtained by combining linearly the statistical and
    slicing errors. The dashed line indicates $\ptcut=4\,\GeV$, as used in~\cite{Camarda:2021ict,Camarda:2021jsw}.}
\end{figure}

We repeat our \ntlo{} calculation using the $\rd\sigma_{\text{DY+jet}}^{\nnlo} $ predictions and setup of Ref.~\cite{Bizon:2019zgf} to compare with Refs.~\cite{Camarda:2021ict,Camarda:2021jsw}. These references employ the same data to extract the \ntlo{} fiducial cross section with symmetric cuts for the central scale choice $\mur = \muf = \sqrt{\mll^2+\pt^2}$.
Following Refs.~\cite{Camarda:2021ict,Camarda:2021jsw}, we perform the calculation using $\ptcut=4\,\GeV$ with and without linear power corrections.
Fig.~\ref{fig:ptcut-suppl} shows that the omission of linear power corrections as done in Ref.~\cite{Camarda:2021ict} leads to a result for the \ntlo correction that is off by ${\cal O}(30\%)$.
These linear power corrections are instead accounted for in Ref.~\cite{Camarda:2021jsw}, although the slicing cut at $\ptcut=4\,\GeV$ is employed already at ${\cal O}(\alpha_s)$. This introduces a systematic slicing error of $\mathcal{O}(\as\, (\ptcut/\mll)^2)$, corresponding to a shift of about $2\,\pb$, not quoted in the uncertainty of Ref.~\cite{Camarda:2021jsw}.

Nevertheless, focusing on the \ntlo{} correction alone, we find that the data of Ref.~\cite{Bizon:2019zgf} is not of sufficient quality to observe a stable plateau around $\ptcut=4\,\GeV$, as can be seen in Fig.~\ref{fig:ptcut-suppl}. When applying the procedure described in the letter for the estimate of the residual uncertainty, we find a statistical error of $1.3\,\pb$ and a slicing error of $1.5\,\pb$ (computed using the range $\ptcut \in [3,5]\,\GeV$). When combined, the total uncertainty is three times larger than the $0.7\,\pb$ that Ref.~\cite{Camarda:2021jsw} quotes.

\subsection{Fit-based implementation of $q_T$ subtraction}
As an alternative to choosing a specific $\ptcut$, a fitting procedure can be employed either by extrapolating Fig.~2 of the main letter down to $\ptcut \to 0$ or by analytically integrating a fit function.

In the first method we directly fit the \ntlo{} curve in Fig.~2 in the range $\ptcut \in [\ptcutmin,\,\ptoff]$ using a parametrisation that follows the known analytic structure of power corrections
\begin{equation}
 \label{eq:cumulant}
\Delta\sigma^{\ntlo}(\ptcut) \sim \Delta\sigma^{\ntlo}_0 +\frac{\as^3}{(2\pi)^3}\,\frac{\ptcut^2}{\mll^2}\, \sum_{n=0}^{5}a^{(3)}_{n}\,\ln^n\frac{m_{\ell\ell}}{\ptcut},
\end{equation}
and extract $\Delta\sigma^{\ntlo}_0$. The lower edge of the fitting range $\ptcutmin$ is set to $0.45\,\GeV$. In the second method, similarly to what has been done in Ref.~\cite{Billis:2021ecs}, we consider a fit up to relatively large $\pt\leq\ptoff$ values of the residual quadratic power corrections $\rd[ \Delta  \sigma_{\text{DY}}^\text{non-sing}]_{\as^3}$, which denotes the $\as^3$ coefficient of the difference $\rd\sigma_{\text{DY+jet}}^{\nnlo} - \big[\rd\sigma_{\text{DY}}^{\ntll}\big]_{{\mathcal{O}(\as^3)}}$.
The fiducial cross section is then obtained by performing a slicing
calculation with $\ptcut=\ptoff$ and adding the analytic integral of
the above fit over $\pt \in [0,\,\ptoff]$.
In this case, the functional form is
\begin{equation}
\label{eq:differential}
\rd[ \Delta  \sigma_{\text{DY}}^\text{non-sing}]_{\as^3} \sim \frac{\as^3}{(2\pi)^3}\,\frac{\pt}{\mll}\, \sum_{n=0}^{5}b^{(3)}_{n}\,\ln^n\frac{m_{\ell\ell}}{\pt},
\end{equation}
with the coefficients $b^{(3)}_{n}$ being extracted from the fit.

We perform Markov chain Monte Carlo fits based on generative probabilistic models for the data that facilitate a straighforward marginalisation over nuisance parameters and the propagation of uncertainties and their correlations.
The latter are relevant for the extrapolation at the cumulant level, as the integral of $\rd\sigma_{\text{DY+jet}}^{\nnlo}$ features large off-diagonal entries in the covariance matrix.
In both fitting methods we find that too small values of $\ptoff$ do not allow for the inclusion of sufficiently many data points to constrain the fit parameters, yielding uncertainties as large as ${\cal O}(100\%)$ for the \ntlo{} correction. On the other hand, choosing larger values of $\ptoff\gtrsim 50\,\GeV$ leads to a more stable determination of the \ntlo correction, which in both cases is in agreement with the result quoted in Tab.~I of the letter within uncertainties.
The nominal uncertainty is slightly smaller than the slicing uncertainty obtained with the procedure described in the letter.
However, increasing the value of $\ptoff$ induces a higher sensitivity to yet subleading-power terms in Eqs.~\eqref{eq:cumulant}--\eqref{eq:differential}, which constitute an additional source of systematic uncertainty that must be reliably assessed.
A naive extension of the fitting functional forms to account for an additional tower of next-to-subleading terms increases the number of fitting parameters, ultimately leading to uncertainties which are significantly larger than those quoted in Tab.~I.
For this reason, we choose to adopt the slicing procedure described in the main letter, which allows for a reliable assessment of all the sources of uncertainty involved.